\documentclass[prl,superscriptaddress,showpacs,showkeys,twocolumn]{revtex4}
\usepackage{graphicx,amssymb,bm}

\newcommand{\Gd}{Gd$^{3+}$}
\newcommand{\Y}{Y$^{3+}$}
\newcommand{\Ca}{Ca$^{2+}$}
\newcommand{\YBCO}[1]{YBa$_2$Cu$_3$O$_{#1}$}
\newcommand{\YCa}{Y$_{1-x}$Ca$_x$Ba$_2$Cu$_3$O$_6$}
\newcommand{\Ms}{\ensuremath{\bm{M}_{\text{s}}}}

\begin{document}

\title{Diagonal Antiferromagnetic Easy Axis in Lightly Hole Doped \YCa}

\author{Andr\'as J\'anossy}
\email{atj@szfki.hu}
\affiliation{Institute of Physics, Budapest University of Technology and
Economics, and Solids in Magnetic Fields Research Group of the Hungarian
Academy of Sciences, P.O. Box 91, H-1521 Budapest, Hungary}

\author{Titusz Feh\'er}
\affiliation{Institute of Physics, Budapest University of Technology and
Economics, and Solids in Magnetic Fields Research Group of the Hungarian
Academy of Sciences, P.O. Box 91, H-1521 Budapest, Hungary}
\affiliation{Institute of Physics of Complex Matter, EPFL, CH-1015
Lausanne, Switzerland}

\author{Andreas Erb}
\affiliation{Walther Meissner Institut, Bayerische Akademie der
Wissenschaften, D-85748 Garching, Germany}

\date{\today}

\begin{abstract}

Hole induced changes in the antiferromagnetic structure of a lightly Ca
doped Gd:\YCa\ copper oxide single crystal with $x\approx0.008$ is
investigated by \Gd\ electron spin resonance. Holes do not localize to
\Ca\ ions above $2.5\,\text{K}$ since the charge distribution and spin
susceptibility next to the \Ca\ are independent of temperature. Both
hole doped and pristine crystals are magnetically twinned with an
external magnetic field dependent antiferromagnetic domain structure.
Unlike the undoped crystal, where the easy magnetic axis is along
$[100]$ at all temperatures, the easy direction in the hole doped
crystal is along the $[110]$ diagonal at low temperatures and changes
gradually to the $[100]$ direction between $10\,\text{K}$ and
$100\,\text{K}$. The transition is tentatively attributed to a magnetic
anisotropy introduced by hole ordering.

\end{abstract}



\maketitle

\begin{figure}
\includegraphics[width=0.8\columnwidth]{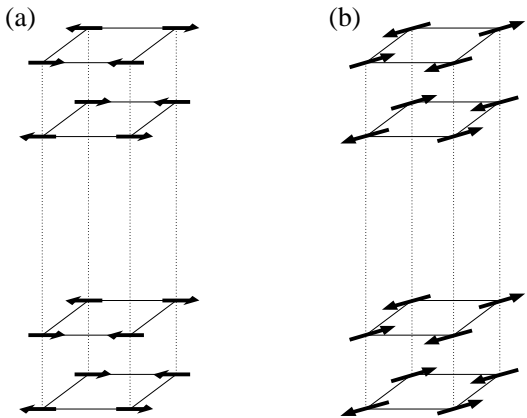}%
\caption{\label{fig:AF-struct}(a) AFI(0) is the magnetic structure of
undoped \YBCO6\ at all temperatures below $T_{\text{N}}=420\,\text{K}$.
(b) AFI($\pi/4$) is the suggested low temperature magnetic structure of
\YCa.}
\end{figure}

The structure of the magnetic elementary cell is well established in
\YBCO6, the antiferromagnetic (AF) parent compound of a high temperature
superconductor, but little is known about how introducing holes affects
the magnetic structure. Neutron diffraction found a chessboard like AF
order in the basal $(a,b)$ plane \cite{Tranquada88} which, in high
purity crystals, alternates along the $c$ direction with the periodicity
of the lattice as shown in Fig.~\ref{fig:AF-struct}(a). In
\YBCO{6+\delta}\ no hole induced change of this structure has been
reported up to date. However, in La$_{2-x}$Sr$_x$CuO$_4$, the other well
studied AF parent compound for high $T_{\text c}$ superconductivity,
hole doping changes the magnetic structure in a peculiar way
\cite{Wakimoto1999,Matsuda2000}. At low temperatures a static magnetic
modulation appears with a wavelength roughly proportional to the hole
concentration and is characterized by a single wavevector running along
one of the diagonals of the slightly distorted square CuO$_2$ lattice.
It is not yet clear whether this spin density wave is or is not a
manifestation of a theoretically predicted \cite{Zaanen,Emery} phase
separation into hole-rich and undoped regions. Tranquada {\it et al.}\
\cite{Tranquada95} suggested that the stripe like charge and magnetic
order in La$_{1.6-x}$Nd$_{0.4}$Sr$_x$CuO$_4$ and related compounds is
evidence for such a phase separation. Ando {\it et al.}\ \cite{Ando2002}
interpreted the doping dependence of the resistivity in terms of
segregated conducting stripes of holes. However, to present, electron
structural studies \cite{Yoshida} do not show the expected quasi 1D
Fermi surface planes at low concentrations.

In this work we observe a hole induced change in the AF structure in a
single crystal of Gd:\YCa\ with a Ca concentration of $x\approx0.008$.
\Ca\ substitutes for \Y\ and introduces 0.004 holes per Cu(2) into the
AF CuO$_2$ layers of our sample. We show that in the lightly hole doped
AF crystal, holes are not localized in the neighborhood of \Ca\ ions
down to $2.5\,\text{K}$. We observe AF domains that are static on the
time scale of $10^{-8}\,\text{s}$ at temperatures below $200\,\text{K}$.
As opposed to La$_{2-x}$Sr$_x$CuO$_4$, \YBCO6\ has a tetragonal crystal
structure and is therefore sensitive to any small extra anisotropy. We
find that the easy axis of sublattice magnetization rotates by
$45^\circ$ as the temperature is lowered. Unlike the undoped reference
that has the structure of Fig.~\ref{fig:AF-struct}(a) at all
temperatures, in the doped sample the ground state spin orientation is
along the diagonal of the CuO$_2$ square lattice
[Fig.~\ref{fig:AF-struct}(b)]. The change of the magnetic anisotropy of
the bulk by a very small concentration of holes is an indication of an
ordered hole structure.

We use 1\% of Gd substituting for Y as a weakly perturbing local
electron spin resonance (ESR) probe of the magnetic spin susceptibility
and charge redistribution in the AF CuO$_2$ sandwich. The \Gd\ ESR $g$
factor shifts and zero field splitting (ZFS) parameters are closely
related to these quantities. Ref.~\cite{jan99} details the methods used
to precisely determine the $g$ factor and the ZFS parameters from the
\Gd\ ESR fine structure lines. This technique has proven successful in
the study of AF \YBCO{6+\delta}\ \cite{jan99,jan90} and La$_2$CuO$_4$
\cite{Rettori} compounds.

Single crystals of Gd:\YCa\ and the reference crystal Gd:\YBCO6\ were
grown in BaZrO$_3$ crucibles as described elsewere \cite{Erb}. The
nominal Gd concentration was 1\% in agreement with estimates from the
intensity of ESR lines from pairs of neighboring \Gd\ ions. The nominal
Ca concentration was $x_{\text{nom}}=0.03$, but due to the small
distribution coefficient of Ca during crystal growth the actual
concentration $x$ determined by atomic absorption spectroscopy (AAS) is
much smaller, between 0.005 and 0.01. A Ca concentration of $x=0.005$ is
estimated from the Ca satellite ESR intensity. We shall refer to the Ca
concentration as $x=0.008$.

\begin{figure}
\includegraphics[width=0.9\columnwidth]{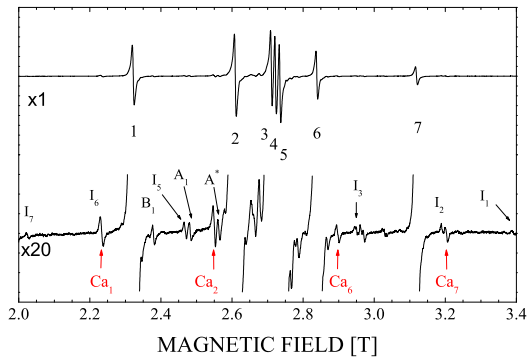}%
\caption{\label{fig:assign}Assignment of lines in the ESR spectrum of
Gd:\YCa. ``1'' to ``7'': the main \Gd\ fine structure series arise from
\Gd\ with no \Ca\ neighbors. ``Ca$_1$'', ``Ca$_2$'', ``Ca$_6$'' and
``Ca$_7$'' are the Ca satellite series from \Gd\ ions with one first
neighbor \Ca\ ion. Other lines are independent of Ca doping and are
described in the text. $B\parallel c$,
$\omega_{\text{L}}/2\pi=75\,\text{GHz}$ and $T=18\,\text{K}$.}
\end{figure}

The ESR spectra testify the high purity of the sample. The powder ESR
spectrum of a small amount of a polycrystalline paramagnetic impurity
phase (possibly the Y$_2$BaCuO$_5$ ``green phase'') and the \Gd\ fine
structure (see Fig.~\ref{fig:assign}) of a low concentration (about
0.5\% of the \Gd\ ions) of an unidentified Gd site within the single
crystal was observed. The X-band ESR spectra of two Ca doped crystals
measured at several orientations at $77\,\text{K}$ were the same and one
crystal of $2{\times}3{\times}0.2\,\text{mm}$ was investigated in
detail. The ESR spectra of the reference samples reproduced those
reported in Ref.~\cite{jan99} except for a somewhat broader linewidth
and a stronger field dependence of the domain structure.

For magnetic field $B$ along the $c$ axis, a single series of strong
fine structure lines (main lines) is observed, corresponding to the
majority of \Gd\ ions with only \Y\ ions occupying the rare earth sites
in the neighborhood (Fig.~\ref{fig:assign}). These lines, labeled ``1''
to ``7'', are well fitted by simulated spectra using the same ZFS
parameters as for the undoped crystal. Several lines with intensities of
the order of 1\% of the main lines appear at larger amplification. Most
of these were observed in undoped samples also and they are due to
closely spaced \Gd-\Gd\ pairs \cite{Simon99} (``A$_1$'', ``A$^*$'' and
``B$_1$'' in Fig.~\ref{fig:assign}). The unidentified Gd site (series
``I$_n$'') has also been observed in previous studies \cite{jan90}.

We assign the satellite lines, ``Ca$_1$'', ``Ca$_2$'', ``Ca$_6$'' and
``Ca$_7$'', to \Gd\ ions which have a first neighbor rare earth site
occupied by a \Ca\ ion. (The inner three lines of the series could not
be identified.) This series has not been observed in the undoped
crystals. The relative intensities, the temperature dependence of the
intensities and the Larmor frequency dependence clearly show that these
lines belong to a \Gd\ fine structure series corresponding to a rare
earth site with ZFS parameters not very different from the main line
series. The strong first neighbor Ca satellite fine structure series of
a magnetically aligned superconducting
Gd:Y$_{0.85}$Ca$_{0.15}$Ba$_2$Cu$_3$O$_{6+\delta}$ powder has been
easily identified previously \cite{Roki-ampere}. The shifts of the
satellites from the main series in the $x=0.15$ sample are similar to
those in the $x=0.008$ sample studied here.

\begin{figure}
\includegraphics[width=0.95\columnwidth]{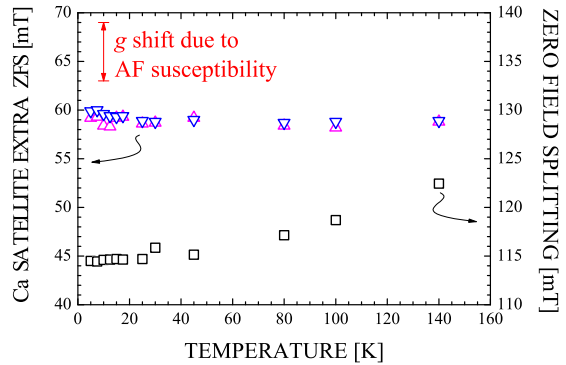}%
\caption{\label{fig:Ca-pos} Left hand scale: ``Ca$_2$''
($\triangledown$) and ``Ca$_6$'' ($\vartriangle$) satellite line
positions measured from the corresponding ``2'' and ``6'' main lines.
The temperature independence of this extra ZFS shows that holes do not
localize at \Ca\ ions. The $g$ shift caused by the spin susceptibility
of the CuO$_2$ layers (marked by vertical arrow) is equal on the
satellite and main lines and is independent of temperature. Right scale:
Half distance between main lines ``2'' and ``6''. The temperature
dependence is due to lattice expansion.}
\end{figure}

We first address the question: do the holes localize at low temperatures
to the Ca sites or elsewhere? We compare the temperature dependence of
the charge distribution and spin susceptibility at sites next to and
distant from \Ca\ ions. \Gd\ ZFS parameters are highly sensitive to the
charge distribution. Doping Gd:\YBCO{6+\delta} with oxygen from
$\delta=0$ to $\delta=1$ changes the largest ZFS parameter, $b_2^0$, by
40\% \cite{Roki92} and changes of this order are expected for the first
neighbor \Gd\ if holes were to localize to the \Ca\ ions within a few
lattice constants. Hole localization would then shift the ``Ca$_n$''
satellite lines on the order of their distances from the central (``4'')
line. This is not the case. The observed small, few \%, temperature
dependence of the ZFS parameters (Fig.~\ref{fig:Ca-pos}, squares) is due
to thermal expansion of the lattice since the temperature variation is
the same at the first neighbors of \Ca\ and at sites distant from the
\Ca\ ions. The extra ZFS caused by the \Ca\ neighbor
(Fig.~\ref{fig:Ca-pos}, triangles), measured by the difference between
the ``Ca$_2$'' (``Ca$_6$'') and ``2'' (``6'') line positions, is
temperature independent within $\pm2$\% from 2.5 to $140\,\text{K}$.
Thus holes do not localize at \Ca\ ions (at least above $2.5\,\text{K}$)
since the charge redistribution at Cu(2) and O(2) lattice atoms
surrounding the \Ca\ ions would inevitably change the ZFS. Also, the $c$
axis susceptibilities (measured by the $g$ factors) near the Ca site and
in the bulk are the same within $\pm10$\% in the temperature range of
2.5 to $140\,\text{K}$ (Fig.~\ref{fig:Ca-pos}) and this reinforces the
conclusion that localization of holes does not take place at Ca sites.
The equal susceptibilities at the satellite and main sites exclude the
possibility that charges are already localized to \Ca\ ions at high
temperatures.

As shown below, even a light hole doping has a dramatic effect on the
magnetic structure. The temperature independent AF domain structure
\cite{Burlet,jan99} of pristine \YBCO6\ was confirmed for the reference
sample. The easy direction of the sublattice magnetization \Ms\ is
along $[100]_t$ (any of the equivalent $[100]$ or $[010]$ directions in
the tetragonal crystal) in the undoped antiferromagnet. In zero applied
field, single crystals are magnetically twinned with equal amounts of
domains in the two orthogonal directions, $[100]_t$, of the tetragonal
lattice. The domain structure is field dependent, and applying a
magnetic field of a few T along a $[100]_t$ direction turns all domains
perpendicular to the applied field. In the present reference sample
about 80\% of domains had $\Ms$ perpendicular to the field in a field of
$1\,\text{T}$ applied along $[100]_t$.

Nothing is known from experiments about the direction, width,
concentration or origin of the domain walls separating the orthogonal
domains. Domain walls in the $(a,b)$ plane that separate domains along
$c$ were suggested to explain data in undoped crystals \cite{jan99}.
Conducting stripes running along $a$ and separating domains within the
$(a,b)$ plane were suggested by Niedermayer {\it et al.}\
\cite{Niedermayer} and Ando {\it et al.}\ \cite{Ando99} to explain NMR
and magnetoresistance data.

The structure of the ESR lines reflects the distribution of domain
orientations. There are two contributions to the \Gd\ resonance
frequency that depend on the orientation of $\Ms$: one formally
described by a ``$g$ shift'' and the other by an orthorhombic ZFS
parameter $b_2^2$. Most probably, both are due to the exchange
interaction $J$ between the CuO$_2$ planes and \Gd\ ions and are
proportional to the anisotropic AF spin susceptibility. $b_2^2$ may also
arise from magnetostriction, but the experiments described here suggest
that a term second order in $J$, overlooked in Refs.\ \cite{jan99} and
\cite{Rettori}, is dominant.

\begin{figure}
\includegraphics[width=0.95\columnwidth]{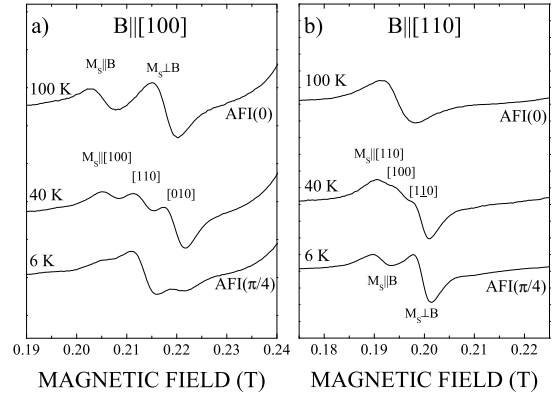}%
\caption{\label{fig:ab-spectra}Magnetic phase transition between AFI(0)
and AFI($\pi/4$) observed in the temperature dependence of a low field
\Gd\ ESR fine structure line at $9\,\text{GHz}$. The direction of
sublattice magnetization \Ms\ at resolved lines is marked. Magnetic
field $\bm{B}$ is applied along (a) $[100]$ and (b) $[110]$.}
\end{figure}

The AF domain structure is readily apparent in the \Gd\ ESR spectra with
magnetic field applied in the $(a,b)$ plane. Fig.~\ref{fig:ab-spectra}
displays one component of the fine structure of the $9\,\text{GHz}$ ESR
spectra at several temperatures for magnetic fields oriented along
$[100]_t$ (i.e., magnetic field $\bm{B}$ oriented along
$\phi_{\bm{B}}=0^\circ$ or $90^\circ$) and $[110]_t$
($\phi_{\bm{B}}=45^\circ$ or $135^\circ$). At high temperatures the
spectra of the Ca doped and undoped \YBCO6\ crystals are similar: the
main series are split by the same amount and the line widths are
comparable. In contrast to the reference, the ESR spectrum of Gd:\YCa\
is temperature dependent. A gradual magnetic phase transition is
observed around $T_m=40\,\text{K}$. Spectra at $T\gg T_m$ and $T\ll T_m$
are qualitatively different. The transition is broad and non trivial
changes with temperature are still observed above $85\,\text{K}$ and
below $15\,\text{K}$. The domain structure and the transition depend on
magnetic field. An account of the transition between 0.2 and
$5.6\,\text{T}$ will be presented elsewhere.

At lower temperatures the magnetic easy axis is rotated by $\pi/4$.
Below about $20\,\text{K}$, the $\phi_{\bm{B}}$ dependence of the lines
is shifted by $\pi/4$, i.e., the lines in the $\bm{B}\parallel[100]_t$
spectrum are similar to the corresponding ones at high temperature in
the $\bm{B}\parallel[110]_t$ spectrum. We denote the high and low
temperature orders as AFI(0) and AFI($\pi/4$), respectively. The
probable cell for AFI($\pi/4$) is shown in Fig.~\ref{fig:AF-struct}(b),
but other, larger magnetic cells are also possible. The high temperature
spectra are well modeled by the same spin Hamiltonian with the same ZFS
parameters as for the undoped case, and this is still true at low
temperatures except for a $\pi/4$ rotation of the orthorhombic term
$b_2^2$.

At high temperatures the system is purely AFI(0) within experimental
uncertainties while it is almost purely AFI($\pi/4$) at the lowest
temperatures. During the transition the system is inhomogeneous but it
is not possible to tell from an analysis of the ESR spectra whether
during the phase transition domains rotate continuously or a mixture of
the two phases occurs. The lack of hysteresis with cycling the magnetic
field and temperature makes a continuous transition more probable.

The AFI(0) to AFI($\pi/4$) phase transition observed here in a high
purity, lightly Ca doped crystal has no relation with the AFI(0) to AFII
phase transition observed in strongly doped crystals where dopants give
rise to magnetic moments in the Cu(1) layer \cite{Kadowaki,Brecht}. In
the AFII phase every second bilayer is rotated by $\pi$ and the lattice
is doubled along $c$. This could be easily distinguished from the
AFI($\pi/4$) state by ESR.

We believe the transition is related to hole localization in spite of
the low hole concentration $p=0.004$. The sample is of high purity and
extrinsic effects are unlikely. The broad temperature range of the
magnetic transition (i.e., a continuous change in the distribution of
the orientations of \Ms\ between 100 and $6\,\text{K}$) is consistent
with the idea that the magnetic transition is related to the
localization of holes. The gradual localization of holes is well
documented in Ca doped \YBCO6 \cite{Niedermayer}. At low
temperatures the sublattice magnetization of the bulk becomes the same
in hole doped systems as in undoped ones, and for Ca concentrations
$x=0.03$ to 0.06 a broad temperature range of magnetic fluctuations and
a magnetic transition to a static order in the range of
10--$20\,\text{K}$ is observed.

Our findings suggest that in \YCa\, in analogy to lightly doped
La$_{2-x}$Sr$_x$CuO$_4$, holes localize into an ordered structure which
induces a diagonal spin density wave with spin direction coupled to the
propagation vector. The amplitude of the spin density wave is small and
the change of the easy axis direction is not accompanied by an
observable change of the magnitude of the magnetic susceptibility.
However, localized holes change both the direction and the magnitude of
the magnetic crystalline anisotropy, since at low temperatures the
domain structure persists to high magnetic fields. In the Ca doped
sample AF domains are well observed at $5.4\,\text{T}$ while they are
hardly discernible in the undoped reference at $2.7\,\text{T}$.

It is probable that the easy axis in pristine \YBCO6\ is determined by
some extrinsic effects, e.g., residual Cu-O chain fragments. This could
explain why no change in the easy direction with temperature was
observed in oxygen doped samples \cite{Casalta}. The difference between
hole doping by Ca and oxygen is that oxygen chain fragments introduce a
local orthorhombic distortion of the structure which at higher
concentrations may fix the direction of magnetic crystalline anisotropy
at all temperatures.

In conclusion the main result is a new magnetic phase transition in
lightly Ca doped \YBCO6\ in which the easy direction of the AF
sublattice magnetization is rotated by $\pi/4$ in the $(a,b)$ plane. We
find that holes do not localize to the Ca dopants suggesting that the
transition is related to the condensation of holes into an ordered
structure. In view of the similarities of the phase diagrams
\cite{Niedermayer}, the diagonal magnetic order in \YCa\ and the
diagonal incommensurate spin modulation observed in doped
La$_x$Sr$_{2-x}$CuO$_4$ may have a common origin.

\begin{acknowledgments}
We are indebted to Patrik Fazekas and Peter Littlewood for illuminating
discussions and to W.~Wendl (Universit\"at Karlsruhe) for the AAS
analysis. Support of the Hungarian state grants OTKA TS040878, T029150,
T043255 and the European Infrastructure Network, SENTINEL, are
acknowledged. The work in Lausanne was partly supported by the NCCR
research pool ``MaNEP'' of the Swiss National Science Foundation.
\end{acknowledgments}

\end{document}